\documentclass[fleqn,12pt]{wlscirep}
\usepackage{setspace}
\usepackage{subcaption}
\usepackage{caption}
\usepackage{color}
\usepackage[symbol]{footmisc}
\usepackage{lineno}
\usepackage{amsmath}
\usepackage{multirow}
\usepackage{adjustbox}
\usepackage{float}
\usepackage{threeparttable}
\usepackage{aas_macros}
\usepackage{txfonts}
\usepackage{ulem}
\title{Quasi-periodic oscillations of GHz-band polarization in a black hole }
\author[1,2,9*] {Wei Wang}
\author[1,2,9] {Jiashi Chen}
\author[1,2,9] {Pengfu Tian}
\author[3,4] {Luis C. Ho}
\author[5] {Xiaohui Sun}
\author[6] {Pei Wang}
\author[7,8] {Bing Zhang}
\author[6] {Zheng Zheng}
\author[1,2] {Xiao Chen}
\author[1,2] {Ping Zhang}
\author[1,2] {Haifan Zhu}
\author[1,2] {Wen Yang}
\author[1,2] {Botao Li}
\affil[1]{Department of Astronomy, School of Physics and Technology, Wuhan University, Wuhan 430072, People's Republic of China;}
\affil[2]{WHU-NAOC Joint Center for Astronomy, Wuhan University, Wuhan 430072, People's Republic of China;}
\affil[3]{Kavli Institute for Astronomy and Astrophysics, Peking University, Beijing 100871, People's Republic of China;}
\affil[4]{Department of Astronomy, School of Physics, Peking University, Beijing 100871, People's Republic of China;}
\affil[5]{School of Physics and Astronomy, Yunan University, Kunming 650500, People's Republic of China;}
\affil[6]{National Astronomical Observatories, Chinese Academy of Sciences, Beijing 100012, People's Republic of China; }
\affil[7] {Nevada Center for Astrophysics, University of Nevada, Las Vegas, NV, USA;}
\affil[8] {Department of Physics and Astronomy, University of Nevada, Las Vegas, NV, USA}

\affil[9] {These authors contributed to this work equally.}
\affil[*] {Email: wangwei2017@whu.edu.cn}

\begin{abstract}
{\bf Relativistic jets from accreting black holes (BHs) radiate non-thermal emission which is highly variable in different time scales \cite{mirabel1999,raiteri2017}. Magnetic fields anchored to a rotating BH \cite{bland1977} or accretion disc \cite{bland1986} accelerate and collimate jets of the BH systems. Previous studies on black holes of different mass scales, including supermassive and stellar-mass black holes, only report flux quasi-periodic oscillations in radio, optical, X-ray and gamma-ray bands \cite{Tian2023,belloni2000,Zhu2024,zhou2018,jorst2022,zhao2024}. No quasi-periodic variations in polarization have yet been detected in any black hole systems. Here, we report the first detection of GHz radio polarization oscillations in GRS 1915+105, which harbors a spinning stellar-mass BH with a relativistic jet \cite{mirabel1994,mcclint2006}. Our observations show that during the increasing phase of radio emission, linear polarization and flux exhibit similar oscillation periods of $\sim 17$ and $33$ seconds, and their variation patterns anti-correlate with each other. These rare, short-period oscillations in both polarization and flux would be important to understand instabilities \cite{dong2020} and special dynamics \cite{ostorero2004,2018MNRAS.474L..81L} in magnetized jets.  }

\end{abstract}

\begin{document}
\maketitle

\begin{figure}
\centering
\includegraphics[width=.6\textwidth]{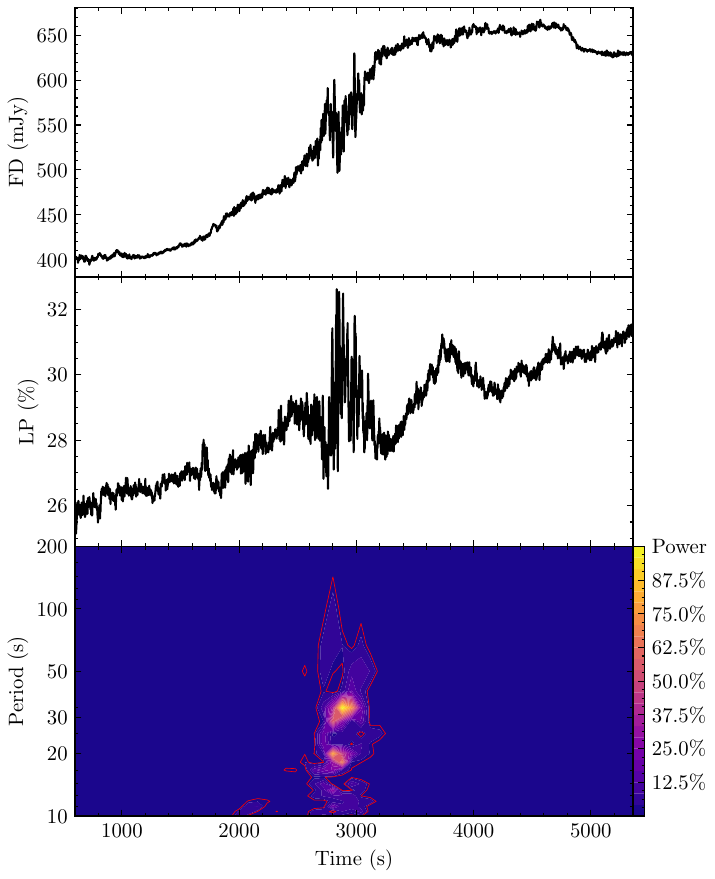}
\caption{{\bf Top panel:} The light curve of total intensity flux density observed by FAST in the 1.05--1.45 GHz band from 2021-01-25:01:45:00 to 2021-01-25:03:04:57 (UT). The flux density began to increase from the time $\sim 1500$ s and reached a peak around 3200 s. {\bf Middle panel:} the light curve of the linear polarization (LP) degree. {\bf Bottom panel:} The dynamical power spectrum of the LP curve for the observed time domain. During the increasing phase of the radio flare, there exist transient quasi-periodic signals at $\sim 17$ and 33 s for LP during the interval  2700--3100 s. The 
orange contour in the represents the line of 95\% confidence levels.}
\label{fig:lc-pds}
\end{figure}

We used the Five-hundred-meter Aperture Spherical radio Telescope (FAST) to perform a high-sensitivity, subsecond-resolution study of GRS 1915+105, aiming to study fast variations and instability of relativistic jets from the BH \cite{tian2023jh}. GRS 1915+105, a well-known Galactic fast-spinning black hole X-ray binary (microquasar) \cite{mcclint2006} at a distance of $\sim$ 8.6 kpc \cite{reid2014}, possesses a relativistic magnetized jet that exhibits superluminal motion through its radio emission\cite{mirabel1994,Fender2004}. GRS 1915+105 was observed by FAST from 01:25:00 - 03:15:00 (UTC) on 25 January 2021 with a 99 microsecond sampling time using the tracking-mode observation pattern on the source in the 1.05-- 1.45 GHz band with the central beam of the 19-beam receiver. At the beginning and end of the observations, two off-source mode observations were performed for calibration. The four full  Stokes polarization parameters were recorded. After data reduction and calibration (Methods), we derived the variations of the total flux intensity and polarization parameters over the observing time intervals with a time resolution of $\sim$ 0.5 s (Figure \ref{fig:lc-pds}). The radio flux density was relatively stable at early times but started to increase after $\sim 1500$ s. After a continuous rise to $\sim 2700$, multiple mini flares appeared for a duration of $\sim 400$ s. The linear polarization degree increases from $\sim 27\%$ to $\sim 31\%$. The bottom panel of Figure \ref{fig:lc-pds} also shows the dynamical power spectrum of the linear polarization light curve, and polarization oscillations with a period of $\sim 17$ and 33 s appear as prominent features during the mini flares in the dynamical power spectrum.

\begin{figure}
\centering
\includegraphics[width=.6\textwidth]{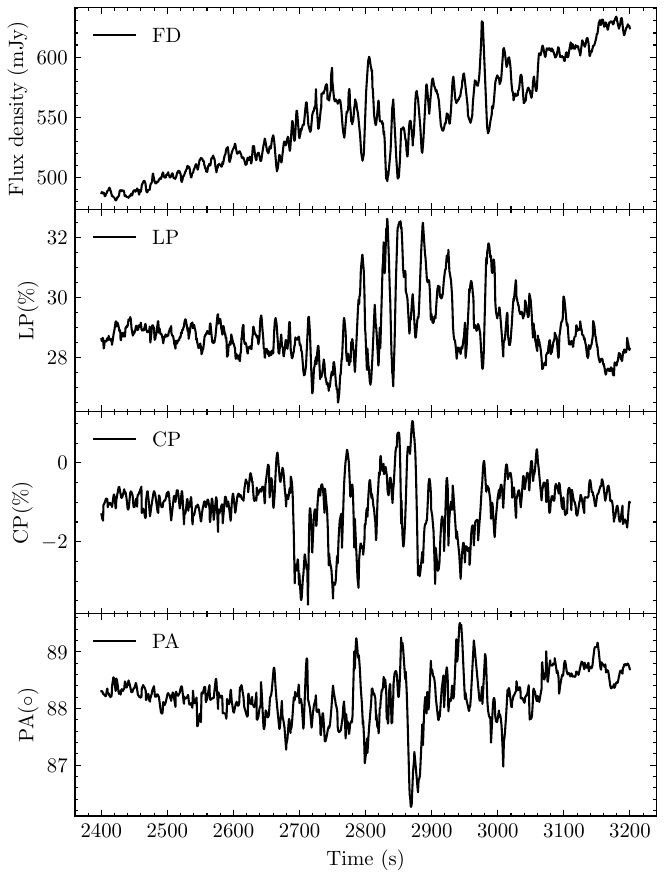}
\caption{The light curves of the flux density (FD), degree of LP, degree of circular polarization (CP), and polarization position angle (PA) from $\sim 2400 -3200$ s when the rapid flux density variations were observed. From $\sim 2700- 3100$ s, the variation pattern of flux density is anti-correlated with that of LP (the peaks of flux density always correspond to the valleys of LP). }
\label{fig:pol_2400}
\end{figure}

\begin{figure}
\centering
% \includegraphics[width=.48\textwidth]{HB-24-1-19-new.pdf}
% \includegraphics[width=.48\textwidth]{HB-24-1-19_LP-n.pdf}
% % \includegraphics[width=.48\textwidth]{HB-24-1-19_CP-n.pdf}
% \includegraphics[width=.48\textwidth]{HB-24-12-30_CP-n.pdf}
% \includegraphics[width=.5\textwidth]{HB-24-1-19_PA-n.pdf}
\includegraphics[width=.47\textwidth]{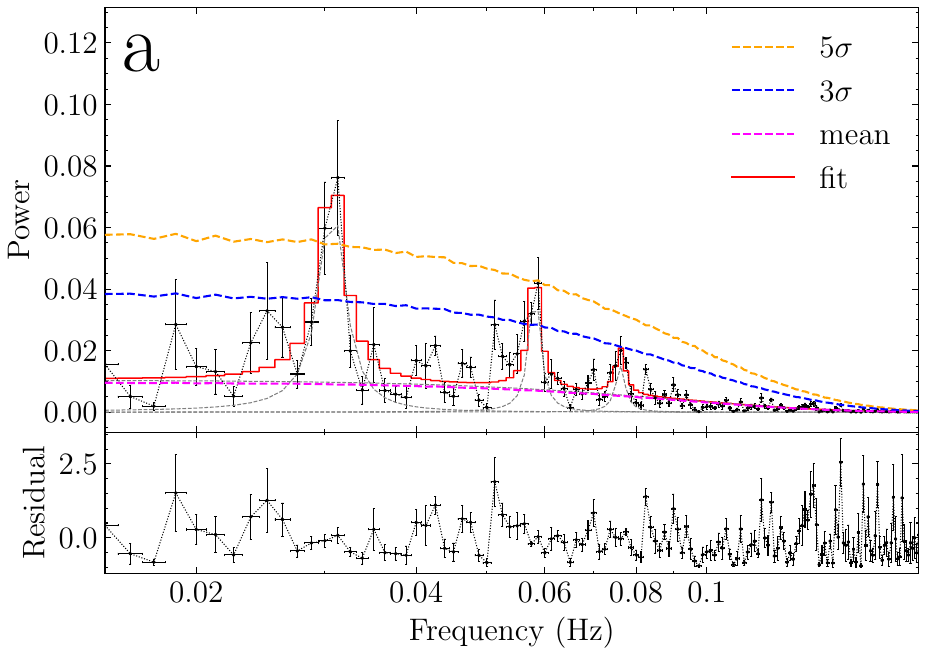}
\hfill
\includegraphics[width=.47\textwidth]{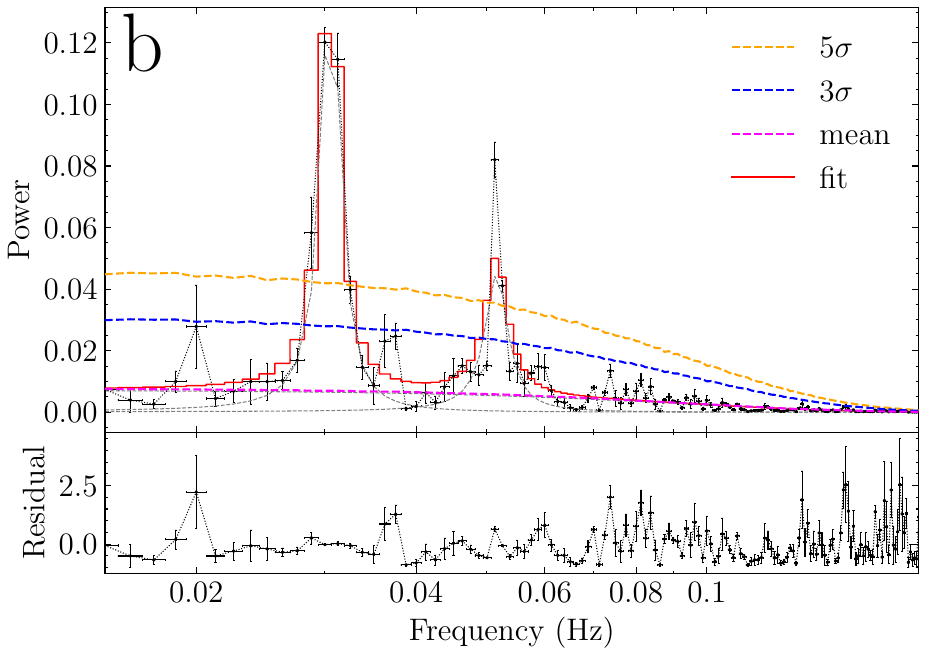}
\vfill
\includegraphics[width=.47\textwidth]{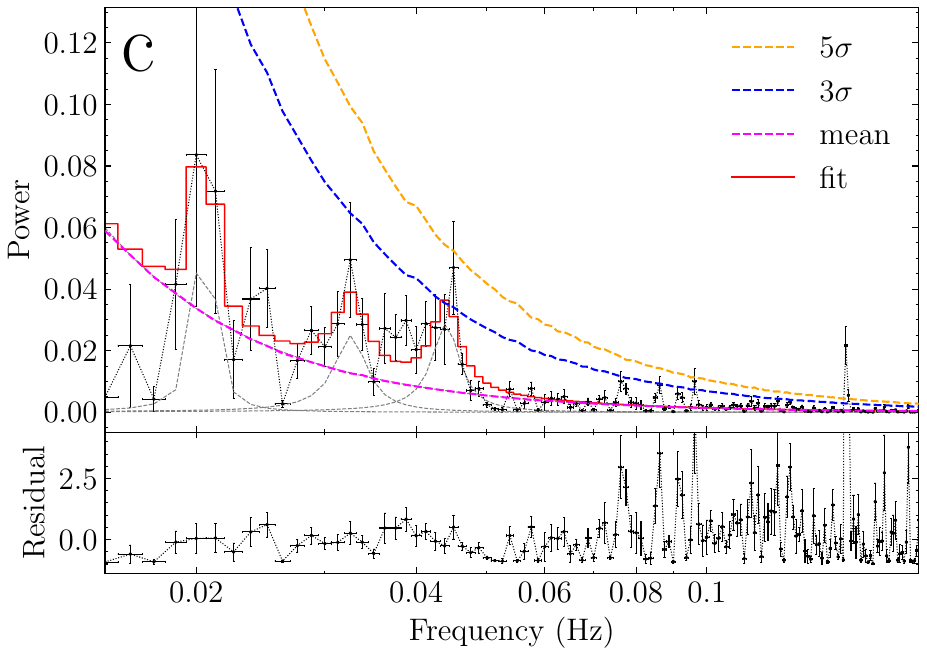}
\hfill
\includegraphics[width=.47\textwidth]{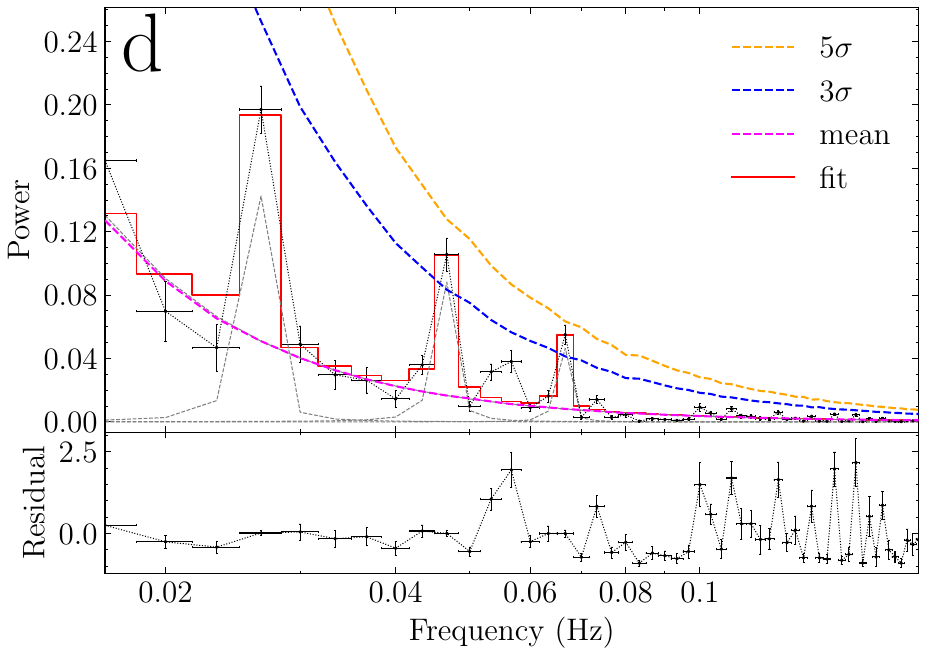}
\caption{The PDSs of the light curves from 2400-3200 s for four parameters: \textbf{(a)} radio flux density (FD), \textbf{(b)} linear polarization (LP), \textbf{(c)} circular polarization (CP) and \textbf{(d)} position angle (PA), and the corresponding residuals of the fittings. The black crosses are normalized with the {\tt rms} normalization, the black dashed lines indicate the Lorentzian components for the fitting, and the red solid line is the result of the fitting. The pink dashed lines indicate the mean background PDSs of the simulated light curves for the four parameters, blue and orange dashed lines indicate the confidence levels of the PDS peaks at 3$\sigma$ and 5$\sigma$ (see Methods). In panel \textbf{a}, the centroid frequencies of the Lorentzian peaks for the FD are around 31, 58 mHz respectively, and there exists a weak peak around 75 mHz which is below $3\sigma$. For LP in panel \textbf{b}, the peak centroid frequencies appear around 30 and 55 mHz above 5$\sigma$, which are similar to the QPO frequencies of the flux. In panel \textbf{c}, the CP light curve shows the different variation patterns, the power has three weak Lorentzian peaks with the centroid frequencies around 20 mHz, 35 mHz and 45 mHz; while in panel \textbf{d}, the PA curve show weak peaks with the frequencies around 25 mHz, 45 mHz and 66 mHz. Both peaks in CP and PA are not significant with  around or below 3$\sigma$.}
\label{fig:pds_2400}
\end{figure}

\begin{figure}
\centering
\includegraphics[width=.66\textwidth]{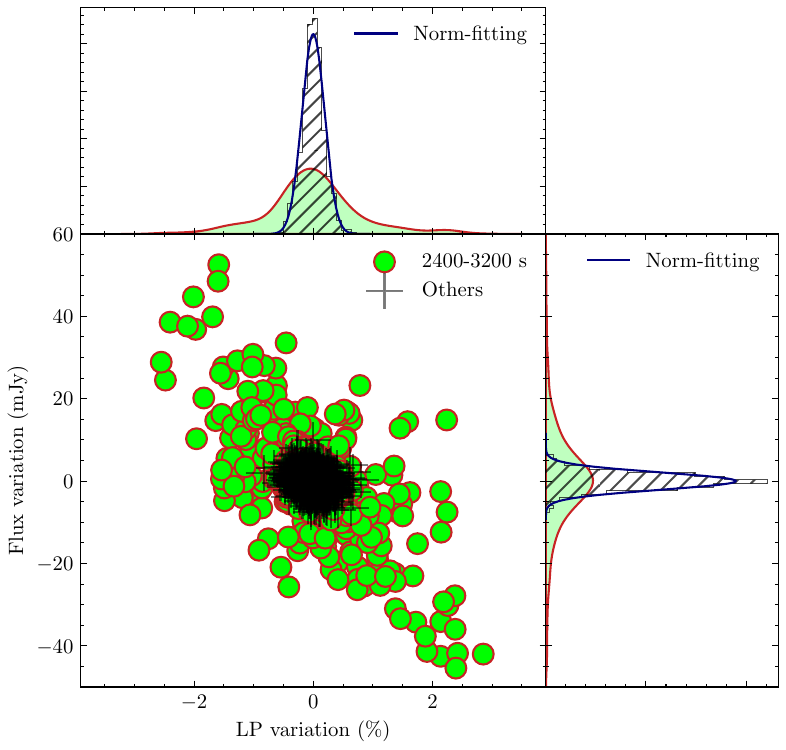}
\caption{Variation patterns between radio flux and linear polarization of GRS 1915+105 based on the FAST observations. During the time interval from 2400 s - 3200 s with the QPO signals, flux and LP show the distribution of the negative correlation,   and the Pearson correlation coefficient (PCC) between flux density and linear polarization during the QPO intervals is -0.795 suggesting a strong anti-correlation. However, for other time intervals without QPO features, the flux and LP distribute randomly and do not correlate with each other. The blue solid lines in the top and right panels are the normal fitting of the histogram.}
\label{fig:flux_lp}
\end{figure}

%After the periodic flares disappeared, the radio flux reached a maximum.
%Faraday rotation measure (RM)
Significant evolution with time for other polarization parameters was also detected during the quasi-periodic flares (see Figure \ref{fig:pol_2400}). We present the detailed variation structure of the flux from $\sim 2400$ to 3200 s, and the corresponding light curves of linear polarization (LP) degree, circular polarization (CP) degree, and position angle (PA). In addition, transient periodic signals with similar periods are confirmed through power density spectral analysis (Figure \ref{fig:pds_2400}). The flux density increased from $\sim 480$ to 620 mJy during this epoch and became dramatically variable between $\sim 2700$ and 3100 s, during which the average power spectrum of the flux density reveals peaks with apparent periods of $\sim 17$ and 33 s. The light curve of the LP degree, which ranges from $\sim (27-32)\%$, exhibits a quasi-periodic modulation (periods around 17 and 33 s) behavior similar to the flux, although the peaks in LP always correspond to valleys in flux: the two are strongly anti-correlated (Pearson correlation coefficient $r=-0.79\pm 0.05$) during the quasi-periodic flares (Figure \ref{fig:flux_lp}). The CP degree varied from $\sim -(0.5 - 2.5)\%$, and the linear PA, distributed at $\sim 88^\circ$, also fluctuated during the quasi-periodic epoch. While the CP and PA curves show the weak transient periodic oscillations at $\sim 40-50$ s, they are also different from those of flux and LP. There are no correlations between the flux and CP or PA (see Methods).

Previous radio observations of GRS 1915+105 have revealed flux quasi-periodic oscillations (QPOs) on timescales of 20--50 minutes\cite{pooley1997,rodri1997,klein2002}. While their physical origin is uncertain, the $\sim$hour timescale of the radio oscillations appears to be connected to X-ray variations on similar timescales \cite{klein2002,fender2002a}. Recently FAST has reported the radio flux oscillations of the period $\sim 0.2$ s \cite{Tian2023}, and this timescale is similar to the low-frequency QPOs with periods of $\sim 0.1-1$ s detected in the X-rays\cite{belloni2000,misra2020}. Here the new findings report, for the first time, rapid QPOs for the light curves of both polarization parameters and flux density in this microquasar with a modulation period down to $\sim 17$ s. In addition, the polarization periodic oscillations only last about twenty periods, in contrast to the relatively persistent low-frequency QPOs in BH X-ray binaries \cite{zhu2024}.

GRS 1915+105 as a bright source in both X-ray and radio bands since its discovery shows complex timing and spectral properties that are quite different from other BH X-ray binaries. The X-ray variability can be classified into approximately 12 separate classes, based on the properties of its light curves and colour-colour diagrams \cite{2000A&A...355..271B,klein2002}, and variability patterns can be further reduced to transitions between three basic states \cite{2000A&A...355..271B}: two soft states and a hard state with different luminosities. In the hard state generally associated with radio jets \cite{2010A&A...524A..29R}, X-ray QPOs with frequencies ranging from 0.1 to 10 Hz are observed \cite{belloni2000,misra2020}. It has been proposed that the radio emission in GRS 1915+105 coming from jets is fed by instabilities in the accretion disk \cite{1997ApJ...488L.109B,1998A&A...330L...9M} by which the inner part of the accretion flow is ejected \cite{2004MNRAS.355.1105F,1998ApJ...494L..61E}. Thus the observed polarized radio QPOs connect to special features of magnetized jets with possible jet-disk couplings.

Several models can explain radio modulations from relativistic jets. It has been suggested that in a rapidly spinning, accreting BH system, a helical magnetic field configuration is naturally produced in the relativistic jets predicted in theoretical magnetohydrodynamical (MHD) models \cite{tchekhov2008,mckinn2014,chen2021}. If the jets are recurrent, the magnetic field may be modulated by some periodic processes \cite{Li2022}, which would lead to modulations of linear polarization as well as the radio flux light curve. The kink instability mechanism, a kind of current-driven plasma instability in magnetically driven jets, can dissipate significant amounts of magnetic energy to accelerate particles\cite{Mizuno2009}. The kink instability can develop in a magnetized relativistic jet with a helical magnetic field configuration. When the jet is disturbed at a height with a displacement, the current-driven kink instability will be triggered \cite{dong2020,Mizuno2009}. Kink instabilities can lead to quasi-periodic energy release that in principle can produce radio flux modulations. Moreover, simulations \cite{dong2020} indicate that the LP degree shows a similar QPO signature and is anti-correlated with the flux.

The radio modulation can also originate from the helical motion of emitting blobs in the relativistic jet \cite{ostorero2004}. The radio and $\gamma$-ray quasi-periodic variability seen on a timescale of tens to hundreds of days in some blazars have been attributed to the helical structure of jet blobs moving \cite{zhou2018,zhang2021}. Magneto-rotational instability (MRI) can lead to a quasi-periodical shock in accretion disk and result in a quasi-periodical flux variation \cite{Okuda2022}, and modulate the magnetic field of jets which may lead to the radio QPO. In addition, disk tearing of a tilted geometrically thin accretion disk due to the Lense-Thirring torque can generate relativistic jets \cite{2018MNRAS.474L..81L,2021MNRAS.507..983L,2023MNRAS.518.1656M}, and these simulations suggest that jets may interact with the outer disk and lead to radio QPOs. However, present simulations on helical motions, MRI or disk tearing have not precisely predicted the polarization behaviors. Therefore, there is no clear evidence that any model is preferred according to the present observations. In any case, the polarized radio QPOs provide the new clues for understanding magnetic field configuration near black holes and particle acceleration in jets. 

\newpage
\section*{Methods}

\section{Observations and data reduction}

We have carried out a FAST observation on the microquasar GRS 1915+105 in tracking mode at 1.05-1.45 GHz band with the central beam of the 19-beam receiver, starting on January 25 2021 01:35:00 (UTC) with a 98 microsecond sample time in a duration of 90 minutes. The resolution of the central beam is $\sim 2.9'$. Before and after the tracking mode observations, the pattern of the on-off mode in which the noise diode was continuously switched on and off was performed for two time intervals: from 01:25:00 - 01:30:00 (UTC) and from 03:10:00 - 03:15:00 (UTC), which are used for calibration processes. The temperature of the reference signal is about 1.1 K for low power mode and 12.5 K for high power mode \cite{jiang2020}, respectively. The measured temperature uncertainty of the diode is $\sim$ 1\%.

The data of FAST are recorded in {\tt PSRFITS} format \cite{hotan2004}. Firstly, we use {\tt astropy} package \cite{astropy2018} to do the preprocessing for {\tt FITS} data files. For each of {\tt FITS} file, we do the re-sampling for the original data and extract the frequency band and time from 4096 frequency channels and 128 subints, and then combine the re-sampled preprocessed data files. The {\tt PRESTO} \cite{ransom2011} would produce a time series which is the uncalibrated lightcurve from the combined file. Meanwhile, {\tt PRESTO} can find the radio frequency interferences (RFI) and create a mask to eliminate these narrow-band signal-like noises. The noisy broadband signal in the periodogram may be caused by RFIs, e.g., the frequency of alternating current in an electronic system may also cause a noise of 50 Hz and the harmonic component in the whole observation time. In addition, the channels from 650 to 820 (1080 MHz -- 1100 MHz) would have the noise and cause many non-negligible short-time spikes (duration of ms) in lightcurves, and meanwhile, there are some RFI broad peaks around channels from 1500 -- 2200, then for a clean data, we cut these channels in the analysis. This time series which has mitigated the RFIs just shows an array of the relative flux intensity to be calibrated in the next.

%We use {\tt rfifind} with a 50-ms time window to find the bad channels. Thus, for the QPOs we detected, the cutting of RFIs found by {\tt rfifind} are sufficiently to restrain the remaining noise level is below the uncertainty of the calibration and ensure the authenticity of the QPOs. To check the performance of {\tt rfifind} to reduce and abandon the noises, we also made a simulation by injecting some frequencies weighted and temporal distributing noises into an off-source data. Then, we use {\tt rfifind} to locate these bad channels and abandon them. The simulation result and test indicate the good performance of {\tt rfifind} picking up RFIs (Fig. \ref{fig:Noisedoffsource}).

{\bf Flux density calibration}

The temperature of the observed source in the on-off mode is
\begin{equation}
\begin{split}
    T_{src}(t) =T_{cal}\cdot\frac{ON}{ONCAL-ON} - T_{sys}(t),
\end{split}
\end{equation}
where $T_{cal}$ is the temperature of the injected reference signal from the noise diode, $ON$ and $ONCAL$ are the intensity values in the turning-off and turning-on states of the noise diode, respectively, when the telescope is directed at the celestial source in the on-off mode for the calibration scans before and after the tracking observations, and $T_{sys}(t)$ is the time-dependent system temperature of the telescope when the telescope is pointing at the background sky, which is given by
\begin{equation}
\label{eq:Tsrc}
\begin{split}
    T_{sys} =T_{cal}\cdot\frac{OFF}{OFFCAL-OFF},
\end{split}
\end{equation}
$OFFCAL$ and $OFF$ are the intensity values in the turning-on and turning-off states of the noise diode, respectively, when the telescope is directed at the background sky.

Then the flux density can be calculated by the following formula,
\begin{equation}
\begin{split}
    Flux(t) = \frac{I(t)-OFF}{ON-OFF}\cdot T_{src}(t)\cdot\frac{1}{G},
\end{split}
\end{equation}
where $I(t)$  is the observed intensity value from the telescope with time in the tracking mode observations, $G=\eta G_0$ differs from the measured gain $G_0$ = 25.6 K/Jy by a factory $\eta$, is the full gain of FAST in the sky coverage, $\eta$ is the aperture efficiency \cite{jiang2020}.

{\bf Polarization calibration} \\
%The original {\tt PSRFITS} files of FAST observation contain the polarization components which are recorded as the AABBCRCI form, where AA and BB are the direct products of two channels, CR and CI are the real and imaginary parts of the cross product of two channels respectively. We use {\tt DSPSR} \cite{straten2011} and {\tt PSRCHIVE}  \cite{hotan2004} to fold the original files with their duration as folding periods to produce the four time series. Each of the four time series is full of 4096 frequency channels and 128 time subints, so it is necessary to eliminate the RFIs manually as the {\tt rfifind} can help us to get the RFIs for each time series.

We use {\tt DSPSR} \cite{straten2011} and {\tt PSRCHIVE}  \cite{hotan2004} to fold the original files with their duration as folding periods to produce the four time series from four recorded channels (stokes parameters), named $I_x^2$, $I_y^2$, CR and CI respectively \cite{straten2010}, where CR and CI are the real and imaginary parts of the cross product of two channels $I_x*I_y$. Generally, for normal devices, feeds are never perfect, and there are two quantities that should be considered and calibrated, i.e., relative gain of an electronic system (leakage) and phase differences between two channels (phase error). And stokes parameters with subscript $obs$ and $true$ refer to polarization components before and after calibration which normally should take the Mueller matrix \cite{heiles2001a,heiles2001b} of the equipment into account respectively, then considering the leakage between two channels, for the linearly polarized signal from the diode, we find
\begin{equation}
\begin{array}{l}
    I_{obs}^{\prime} = I_{true}^{\prime} + f*Q_{true}^{\prime}\\
    Q_{obs}^{\prime} = Q_{true}^{\prime} + f*I_{true}^{\prime},\\
\end{array}
\end{equation}
where $\prime$ means the injected reference signal, and the leakage $f=\frac{Q_{obs}^{\prime}}{I_{obs}^{\prime}}$.
We calibrated the phase error as $\delta_{er}=\frac{1}{2}\arctan\frac{V_{obs}^{\prime}}{U_{obs}^{\prime}}$, then removing the error from orientation, we can get the true values of the four stokes parameters:
\begin{equation}
\begin{aligned}
    I_{true}&=\frac{I_{obs}-f*Q_{obs}}{1-f^2}\\
    Q_{true}&=\frac{Q_{obs}-f*I_{obs}}{1-f^2}\\
    U_{true}&=P\cos[2(\delta-\delta_{er})]\\
            &=P\cos2\delta \cos2\delta_{er}
             +P\sin2\delta \sin2\delta_{er}\\
    V_{true}&=P\sin[2(\delta-\delta_{er})]\\
            &=P\sin2\delta \cos2\delta_{er}
             +P\cos2\delta \sin2\delta_{er},
\end{aligned}
\end{equation}
where $P=\sqrt{U_{obs}^2+V_{obs}^2}$. Finally, the degrees of linear and circular polarization, and polarization position angle are calculated as
\begin{equation}
\begin{aligned}
    LP&=\frac{L}{I_{true}}=\frac{\sqrt{Q_{true}^2+U_{true}^2}}{I}\\
    CP&=\frac{V_{true}}{I_{true}}\\
    PA&=\frac{1}{2}\arctan\frac{U_{true}}{Q_{true}}.
\end{aligned}
\end{equation}

\section{Periodicity studies}

The main aim of the timing analysis here is to search for the quasi-periodic oscillations (QPOs) ranging from $\sim 1- 100$ seconds in the light curves of both radio flux density and polarization parameters. QPOs are generally studied in the Fourier domain and show up in the power density spectrum as narrow peaks.
We used {\tt Numpy.fft.fft} and {\tt Stingray} in {\tt Python} packages to perform the power density spectrum (PDS) analysis, including the production and fitting of the PDS.

The PDS calculated with the packages mentioned above is the Fourier transform of the light curve and is unnormalized. In X-ray observations, such PDS need to be normalized so that the evolution of the QPO can be learned by computing some physical quantities such as the QPO factor for observations of the same QPO but in different equipment. However, unlike the photon number distribution in X-ray observations, the flux density observed in the radio band is highly related to instrument response, system noise, and so on. In order to facilitate the adjustment of fitting parameters and the generation of dynamic PDS, the following procedures are performed: when computing the PDS, we treat the flux density in the radio band as the number of photons in X-ray observations, and then normalize the PDS with the {\tt leahy} and {\tt rms} normalizations. When computing the dynamic PDS, we divide the light curve into segments of 200-second duration and calculate the PDS independently for each segment. To minimize the inaccuracy caused by such crude segmentation and the damage to the light curve structures, each segment is shifted in 80-second steps.

Wavelet analysis is a valuable approach for analyzing time series having a wide range of timescales or variance variations which would decompose the time series into time-frequency space so that we can check both the dominant modes of variability and how those modes vary in time \cite{torr1998}. In reality, wavelet analysis would also perform the FFT of time series; however, unlike traditional FFT for PDS, wavelet analysis would use wavelet functions with different time and amplitude scales for further analysis; however, the most important feature of wavelet analysis is that it would assume a background noise at different scales, which would be useful for calculating the 95\% confidence contour. Meanwhile, the shorts of dynamic PDS which are inevitable would not appear in wavelet analysis. But the mother wavelet utilized in analyzing processing might have a considerable impact on the end outcome, such as how 'Morlet' provides better temporal localization whereas 'Paul' provides better frequency localization. Wavelet analysis, however, has been applied to the timing analysis of light curves at X-ray band \cite{chen2022,ding2021}.

\begin{figure}
\centering
\includegraphics[width=.7\textwidth]{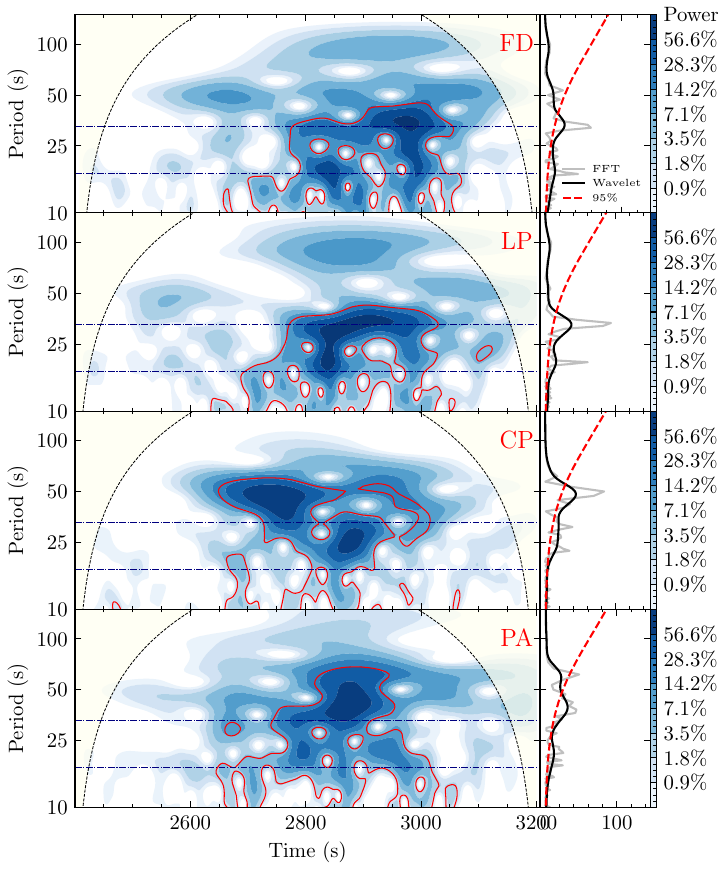}
\caption{Wavelet analysis of the light curves of flux density and polarization parameters (from top to bottom): FD, LP, CP, and PA. The solid red lines indicate the 95$\%$ confidence contours for a red-noise process. Light yellow regions and the black dashed lines indicate the ``cone of influence'' where edge effects become nonnegligible. The two blue dash-dotted horizontal lines in each panel flag the locations of the two QPO signals at 17 and 33 s detected in the linear polarization and flux density light curves. All light curves exhibit oscillations in some time intervals. However, only the LP oscillation signal correlates with the flux density. No significant relationship exists between the other polarization parameters (CP and PA) and flux density. The gray, black solid lines and red dashed lines in the right panels indicate the Fast Fourier transform (FFT) spectra, wavelet spectra and the $95\%$ confidence levels respectively.}
\label{fig:wavelet}
\end{figure}

For wavelet analysis in our data, we have used the 'Morelet' wavelet function for analysis and taken red noise into account in wavelet analysis by calculating the correlation functions of the time series, and a simple model to compute red noise is the univariate lag-1 autoregressive process, so that we estimate the red noise from $(\alpha_1 + \sqrt{\alpha_2})/2$, where $\alpha_1$ and $\alpha_2$ are the lag-1 and lag-2 autocorrelations of the time series. On the other hand, the wavelet analysis would also give the significance levels of the power spectrum based on the red noise model \cite{torr1998}. In our results (such as Fig. \ref{fig:wavelet}), the regions surrounded by black solid curves and colored with deep blue denote the promising signals whose confidence levels are higher than 95$\%$. The result of wavelet analysis would clearly show the evolution of periodic signals both in the time and frequency domains.

At first, we calculated the power density spectrum for every two hundred seconds data set of the calibrated flux time series, checking the possible periodic signals in the PDS, then we arranged them in chronological order to get a set of spectra which is called the dynamical power spectrum (see Fig. \ref{fig:lc-pds} for the whole observational time series on January 25 2021). For the most observational time regimes, there are no periodic signals detected in the dynamical PDS, while only for the time interval from $\sim 2700 - 3100$ seconds when the flux density was still increasing near to the peak, there existed the multiple periodic signals at $\sim$ 17 and 33 seconds. We also presented the average power density spectrum of the calibrated flux density light curve (see the power spectrum in Fig. \ref{fig:pds_2400}) from the observed time interval from 2400 s to 3200 s. The periodic peaks around 17 and 33 s have significance levels higher than 5$\sigma$ based on the light curve simulation algorithm \cite{timme1995}, in which we have simulated 20000 light curves with power-law distributed noises appropriate for our data and re-sampled these light curves to ensure the resolution that matched our observation data.

In Fig. \ref{fig:pds_2400}, we also present the PDSs for the light cures of three polarization parameters (LP, CP and PA). The LP light curve shows similar variation patterns of the flux density and has significant QPO signals at the periods of $\sim 17$ and 33 s ($>5 \sigma$). The CP and PA light curves exhibit very weak QPO signals, with periods of approximately 50 s, 30 s, and 20 s for CP, and 40 s, 30 s, and 15 s for PA. All possible PDS peaks detected in both CP and PA have low significance levels (below or around 3$\sigma$). Therefore, the light curve of LP shows similar variation patterns with that of flux density. The CP and PA curves would have similar variation patterns, but are quite different from the curves of both flux density and LP.

Then, wavelet analysis for flux and polarization parameters has also been conducted. The results are shown in Fig. \ref{fig:wavelet}, in which the light-yellow regions and the red dashed curves indicate the 'cone of influence' caused by the edge effects, the deeply blue colored regions and embraced with red solid curves suggest the existence of promising signals with their confidence levels are higher than 95 $\%$. Meanwhile, the blue dash-dotted horizontal lines in Fig. \ref{fig:wavelet} represent the positions of the two periodic signals for flux and LP light curves ($\sim 17$ and 33 s) in the wavelet analysis results. For a more profound comprehension of the wavelet results as shown in Fig. \ref{fig:wavelet}, one may regard Fig. \ref{fig:wavelet} as results of dynamic PDS calculated by FFT, with abscissa and ordinate indicating the time and period (frequency) domains. The reason of selecting wavelet for timing analysis stems from its ability to dynamically adapt the scale of the mother wavelet according to the local features of the signal. In contrast to traditional FFT, the wavelet analysis method can effectively avoid the negative impact of signal discontinuity caused by truncation, enabling a more accurate and flexible reflection depiction of the signal's instantaneous features and abrupt changes. In our wavelet results, the high time-resolution power density with time shows more details of the variations for these periodic oscillation signals. The strongest period modulation of 33 seconds appeared near 2760 s in both total flux density and LP light curves, but could not be seen at 3050 s. The signal at 17 seconds is relatively weak and has similar time intervals to the 33-second signal. The light curve of CP shows the strong periodic signal at $\sim 50$ seconds from 2600 s to 2800 s, while has a possible 25-second period from 2800 s to 2900 s. And PA also shows the periodic signal around 40-50 seconds from 2800 s to 2900 s.

\section{Correlation analysis}

The light curves in the Figure \ref{fig:pol_2400} clearly illustrate an anti-correlated relationship between flux density and linear polarization during the oscillation stage: the peaks of the flux always correspond to the valleys of the LP. We use linear regression analysis on flux density and linear polarization to confirm this anti-correlation.

Pearson correlation coefficient (PCC) is the best method for measuring the magnitude of the linear association between variables since it is based on the utilization of covariance and is defined as:
\begin{equation}
\begin{split}
    \rho_{X,Y}=\frac{cov(X,Y)}{\sigma_X\sigma_Y},
\end{split}
\end{equation}
where $\rho_{X,Y}$ is the PCC and on or between $-$1 and $+$1 and $\sigma_X$ and $\sigma_Y$ are the standard deviations of $X$ and $Y$ respectively. Since $cov(X,Y)$ is the covariance of variable $X$ and $Y$, so the formula of PCC can be written as:
\begin{equation}
\begin{split}
    \rho_{X,Y}=\frac{\mathbb{E}[(X-\mu_X)(Y-\mu_Y)]}{\sigma_X\sigma_Y}
\end{split}
\end{equation}
We used the {\tt Seaborn} and {\tt SciPy.stats} in {\tt Python} package to do the linear regression analysis and the PCC calculation. The PCC between flux density and linear polarization was then calculated to be -0.795, indicating that flux density has a high anti-linear correlation with linear polarization from the duration of $\sim 2700- 3100$ s when the flux and LP have similar QPO periods. Meanwhile, the PCCs between the flux density and other polarization parameters were also determined, which are smaller than $|-0.25|$ or even smaller, denoting the weak or no relation between these parameters. In addition, before 2700 s or after 3100 s, the variations of the flux and LP have no significant correlation with each other (e.g., PCC $<|-0.2|$) when they show no any periodic oscillations (see Fig. \ref{fig:flux_lp}).

\begin{figure}
\centering
\includegraphics[width=.5\textwidth]{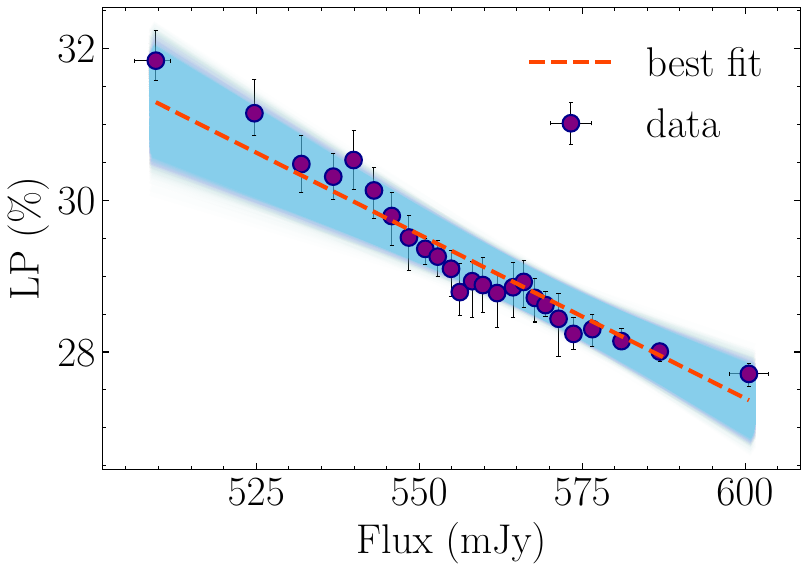}
\includegraphics[width=.5\textwidth]{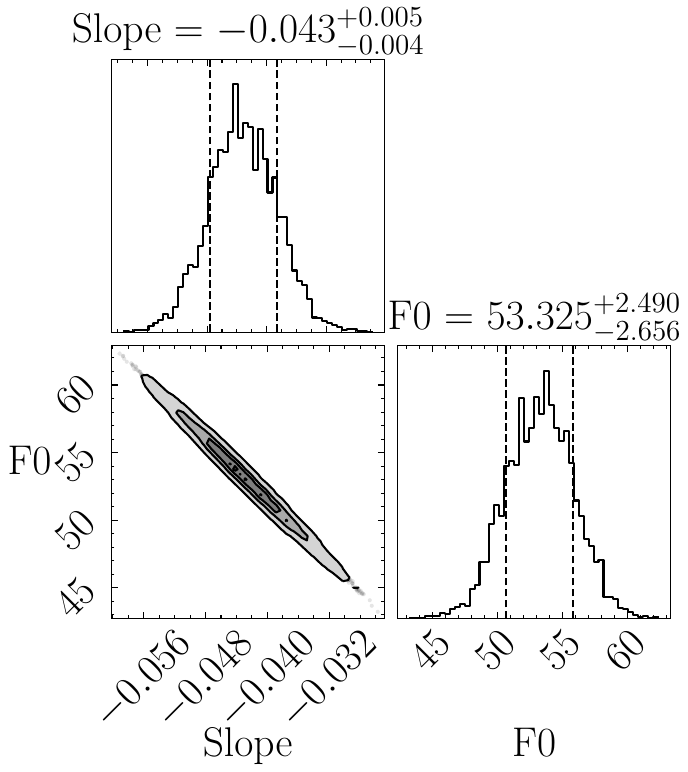}
\caption{{\bf Top panel:} the linear fitting of the linear regression result between flux density and linear polarization for the observed time interval from 2400 s to 3200 s. The purple circles with error bars mean the original scatter points based on flux density and linear polarization. The orange dashed line indicates the best fitting result. {\bf Bottom panel:} the corner plot of the parameters of slope and intercept of the linear fitting shown in the top panel. The fitting was implemented with Markov Chain Monte Carlo (MCMC) sampling. 
}
\label{fig:corr_mcmc}
\end{figure}

In addition, we performed the linear fitting for the relationship between the flux and LP, as shown in Fig. \ref{fig:corr_mcmc}. An apparent anti-correlated relationship is observed between flux density and LP, with the Pearson correlation coefficient of -0.795, and a slope of -0.04.
\section{Possible physical models}
Quasi-periodic modulations in radio light curves, as detected in GRS 1915+105, are frequently observed in different BH systems. Long-period radio oscillations with periods from about one hundred days to several years have been reported in some radio-loud active galactic nuclei (AGNs), specially blazars \cite{zhang2021,Ren2021,raiter2001,Bhatta2017}. These radio oscillations generally last for about several to twenty cycles, which likely reflect the special dynamics of relativistic jets powered by supermassive black holes (SMBHs) in AGNs. Radio oscillations with a period of $\sim$ 15 hours were also found in a gamma-ray X-ray binary LS I$+61^\circ$303 \cite{Jaron2017}, which only had two or three cycles. FAST observations reported the discovery of the transient 5-Hz quasi-periodic oscillations in GRS 1915+105 \cite{Tian2023}. In the previous observations for both stellar mass BHs and SMBHs, however, only the flux modulations are reported, without polarization detection or discovering the periodic oscillations in polarization curves. In this work, the first detection of polarization quasi-periodic oscillations is reported in BH systems.

Many models are proposed to explain quasi-periodic oscillations in radio bands. For stellar mass BH systems, e.g., GRS 1915+105, the half-hour radio periodic oscillations may be connected to the X-ray oscillations with similar periods \cite{klein2002}, while in LS I$+61^\circ$303, it was suggested that the radio modulations could result from multiple shocks in a jet \cite{Jaron2017}. The 5-Hz radio quasi-periodic oscillations in GRS 1915+105 have no good models for the physical origin yet, which could be due to the possible jet precession, jet wobbling, or the helical motion of the relativistic jet knots \cite{Tian2023}. In the framework of AGNs, radio oscillation models are diverse. The year-long oscillations are generally considered to be the indicator of the orbital motion of binary SMBH systems. Helical structures in magnetic fields and plasma trajectory are expected in magnetically dominated jets \cite{chen2021}, so helical motion of blobs or shocks in relativistic jets have been incorporated to interpret periods around hundreds of days in radio, optical, or gamma-ray bands in blazars \cite{zhou2018,Sarkar2021}.

We now consider some of the physical models that may be able to explain the observed radio polarization oscillations in GRS 1915+105. These models would be directly connected to jet dynamics and emission, or arise from disk-jet couplings near the BH. 
\begin{itemize}
    \item[1)] The linear polarization mainly comes from synchrotron emission of electrons in relativistic jets, the polarization degree may be quasi-periodic if the modulation of the magnetic field is quasi-periodic. A group of recurrent jets with similar periods can probably excite the quasi-periodic oscillations in light curves. The recurrent jets may be associated with the magnetic reconnection that is modulated by some periodic processes \cite{Li2022}. The modulated magnetic field associated with the jets could also lead to modulations of linear polarization curves with a similar period to the light curve. However, the origin of recurrent jets is still unclear, and it is unclear whether a group of jets can be generated with a similar interval.
    \item[2)] MHD simulation shows that magneto-rotational instability (MRI) can lead to a quasi-periodical shock in accretion disk and results in a quasi-periodical flux variation \cite{Okuda2022}. The MRI can generate jets and modulate the magnetic field which may lead to the QPO phenomenon. Therefore, the MRI model may explain both X-ray QPOs and radio QPOs observed in hard states of GRS 1915+105. Unfortunately, the present MHD simulations didn't consider polarization properties, so it's unclear how the flux and polarization variation patterns evolve. 
    \item[3)] A model based on helical trajectories of emitting blobs may explain the QPO phenomenon and anti-correlations between the electric polarization angle, the degree of polarization and optical flux \cite{Mangalam2018}. If the QPO originated from helical motion, we should observe the QPO during the whole period of the jet. The QPO only appears during 2700$\sim$3200 s and disappears later which is more likely due to oscillation generated by instability. This geometric origin would be difficult to explain the disappearance of QPOs.
    \item[4)] Recent 3D general relativistic magnetohydrodynamic simulations show that a tilted geometrically thin accretion disk can be torn due to the Lense-Thirring torque \cite{2018MNRAS.474L..81L,2021MNRAS.507..983L,2023MNRAS.518.1656M}. QPOs can be observed due to the precession of infalling accretion matter. In addition, a pair of relativistic jets can be produced in such a system. In some situations, the disk-jet interaction can be very strong that the jets run into the outer sub-disk \cite{2021MNRAS.507..983L}. This can possibly lead to instability in jets and generate QPOs in the radio band. However, the simulations assume a high tilt angle, 45$^{\circ}$-65$^{\circ}$, and there is no clear observation result indicate that GRS 1915+105 has such a high tilt angle.
    \item[5)] Kink instability is a kind of current-driven plasma instability, which causes transverse displacements of plasma and twists the magnetic field structure. Global MHD simulations show that jets will expand and loose transverse causal contact after being launched, making them stable for current-driven instabilities. As the jet moves away from the center black hole, the pressure of the interstellar medium becomes important, and the flow will recollimate and regain its causal contact \cite{Alves2018,Bromberg2019}. Consequentially, the jet becomes narrower and the cross-section is reduced, such area is natural for kink instabilities to develop \cite{dong2020,Rodolfo2017}. The kink instability can distort the magnetic field and generate an induced electric field. The combination of an induced electric field and a distorted magnetic field is an effective nonthermal particle accelerator. Depending on the set up of simulation, $10-50\%$ magnetic energy can be dissipated, which will be transferred to accelerated particles with a power-law spectrum \cite{Alves2018,Davelaar2020}. Particles accelerated via kink instability can give rise to synchrotron emission which accounts for observed polarization signatures \cite{Zhang2017}. Besides, relativistic magnetohydrodynamic simulations show that kinked structure appears with periodic signature along the jet direction, which will lead to quasi-periodic oscillation (QPO) signatures of polarization\cite{Zhang2017,dong2020}. However, the simulation shows the QPO signatures only last for several periods and it's not clear whether the instability can last long enough to generate the observed clear QPOs that last for tens of periods.
\end{itemize}

Jets are prone to many instabilities such as pressure-driven (PD), current-driven (CD), and Kelvin-Helmholtz (KH) instabilities (plus any combination of them). All these instabilities can either trigger radially localized, internal, and surface modes or long-wavelength body modes. The final nonlinear outcome of these instabilities can reach from a simple internal redistribution of jet quantities to the production of internal shocks and/or MHD turbulence. Jets are highly inhomogeneous media, with tremendous magnetic and velocity gradients whose effects on the development of instabilities are not yet fully understood. Nevertheless, there are some models that may explain the radio QPOs in GRS 1915+105. The kink instability model can generate the QPO signal and lead to the anti-correlated relationship between flux density and linear polarization \cite{dong2020}. Several other possible models are also introduced above, which are disk-tearing, magneto-rotational instability, helical moving emitting blobs, and recurrent jets. Unfortunately, so far, there is no clear evidence of which model is preferred according to the present simulations and observation results.

%\begin{figure}
%    \centering
%    \includegraphics[width=\textwidth]{kink1.png}
%    \caption{Sketch of kink instability of the magnetized relativistic jet in the black hole system. %{\bf Left panel:} The helical magnetic field structure of the relativistic jet above the rotating %black hole system is expected. In the region far away from the BH, the instability would develop in %the kinked area of the magnetized jet. {\bf Right panel:} Kink instability will cause transverse %displacements of plasma and twist the magnetic field structure, then dissipate significant amount of %magnetic energy and accelerate non-thermal particles. Kink instabilities cause the quasi-periodic %magnetic energy conversion to thermal energy, which leads to quasi-periodic emission signature.}
%    \label{kink1}
%\end{figure}

%\begin{figure}
%    \centering
%    \includegraphics[width=\textwidth]{mcmc_2.jpg}
%    \caption{Triangle plots of posterior distributions of model parameters for kink instability in the %relativistic jet of GRS 1915+105.}
%    \label{kink2}
%\end{figure}

\section*{Data availability statement}
All FAST data are available from the FAST user website, \url{http://fast.bao.ac.cn}.

\section*{Code availability}
%The data reduction is done by the use of the following software packages.\\
PSRCHIVE (\url{http://psrchive.sourceforge.net})\\
DSPSR (\url{http://dspsr.sourceforge.net})\\
PRESTO (\url{https://github.com/scottransom/presto})\\

\section*{Acknowledgements}
This work is supported by the National Key Research and Development Program of China (Grants No. 2021YFA0718503 and 2023YFA1607901), and National Science Foundation of China (Grants No. 12133007), and the Cultivation Project for FAST Scientific Payoff and Research Achievement of CAMS-CAS. L.C.H. is also supported by the National Science Foundation of China (11991052, 12233001), the National Key R\&D Program of China (2022YFF0503401), and the China Manned Space Project (CMS-CSST-2021-A04, CMS-CSST-2021-A06).

\section*{Author contributions statement}
W.W. as the PI of the FAST observations proposed the project, led the data analysis and wrote the paper. J.C. made the statistical work and discussed related models, P.T. made the data analysis, P.W., X.S., Z.Z., X.C., P.Z., H.Z., W.Y. and B.L. provided the help of the radio data analysis and software. W.W., L.C.H., B.Z. constructed the scientific interpretation to the data. All authors have reviewed the present results and manuscript.

\section*{Competing interests}
The authors declare no competing interests.

{}

\end{document}